\documentclass[english,apj]{emulateapj}

\usepackage[latin9]{inputenc}
\usepackage{graphicx}
\usepackage{amssymb}
\usepackage{esint}
\usepackage{color}

\makeatletter

\shorttitle{Black Hole Triple Dynamics}\shortauthors{Antonini, Murray \& Mikkola}

\makeatother

\usepackage{babel}

\begin{document}

\title{Black hole triple dynamics:  breakdown of the orbit average approximation and implications for gravitational wave detections}

\author{Fabio Antonini$^a$,
  Norman Murray$^a$ \&
  Seppo Mikkola$^b$\\ ~ \\
  $^a$Canadian Institute for Theoretical Astrophysics, University
of Toronto, 60 George St., Toronto, Ontario M5S 3H8, Canada\\
  $^b$Tuorla Observatory, University of Turku, Väisäläntie 20, Piikki\"{o} FI-21500, Finland \\~\\
  E-mail: {\rm antonini@cita.utoronto.ca} }                    

\global\long\def\rsun{{\rm \, R_{\odot}}}
\global\long\def\msun{{\rm \, M_{\odot}}}
\global\long\def\kms{{\rm \, km\, s^{-1}}}
\global\long\def\au{{\rm \, au}}
\global\long\def\mpc{{\rm \, mpc}}

\begin{abstract} 
Coalescing black hole~(BH) binaries forming via dynamical  interactions in the dense core of
globular clusters~(GCs) are expected to be one the brightest and most numerous sources of gravitational 
wave~(GW) radiation, detectable by the upcoming generation of ground based laser interferometers. 
Favorable conditions for  merger are initiated by the Kozai resonance in which the gravitational  interaction 
with a third distant object,  typically another BH, induces quasi-periodic variations  
of the inner BH binary eccentricity.  In this paper we perform high precision  $3$-body 
simulations of the long term evolution of hierarchical BH triples  and investigate the 
conditions that lead to the merging of the BH binary and the way it might become an observable source of GW radiation. We find that the secular orbit average treatment, adopted in previous works,
 does not reliably  describe the dynamics  of these  systems if  the binary is orbited by the outer 
 BH on a highly inclined orbit with a periapsis distance  less than $\sim 10$ times the inner binary semi-major axis.
  During the high eccentricity phase of a Kozai cycle the torque due to the outer BH can drive the binary to extremely 
  large eccentricities in a  fraction of the binary's orbital period. This occurs before relativistic  terms  become important 
  to the evolution and allows the binary  GW signal to  reach large GW frequencies~($\gtrsim 10~$Hz) at high  eccentricities.
We show that $30-50\%$ of coalescing  BH binaries driven by the Kozai mechanism in GCs will have  
eccentricities larger than $0.1$,  with $10~\%$ of them being extremely 
eccentric, $(1-e)\lesssim 10^{-5}$, when they  first chirp in the frequency band of ground 
based laser interferometers.  This implies that  a large fraction of such GW sources could be 
missed  if conventional  quasi-circular templates  are used for analysis of GW detectors data.  
The efficient detection of all coalescing BH binaries in GCs will therefore require  template
 banks of \emph{eccentric}  inspiral waveforms for matched-filtering and dedicated search strategies. 
We conclude by noting  that our results  have potential application  to a variety of  astrophysical 
systems, including  type Ia supernovae, X-ray binaries, optical transients  and their merger products.
\end{abstract}

\section{Introduction}
Coalescence of compact binaries is accompanied by the emission of gravitational waves~(GWs)
at frequencies that are potentially detectable by 
the next generation of ground-based laser interferometers.
Indeed, the first GW signature from a compact binary coalescence could be detected as early as 2015 when
the advanced version of the GW observatories advanced-LIGO~(aLIGO)~\footnote{http://www.ligo.caltech.edu} 
and advanced-VIRGO\footnote{https://wwwcascina.virgo.infn.it/advirgo/}
 will become  operative~\citep{acernese+08,harry+10}.
The aLIGO detectors will be able to see inspiraling neutron stars~(NSs) 
up to a horizon distance of about 400 Mpc, NS-black hole~(BH) binaries will be visible to about 900 Mpc,
and coalescing BH binary systems will be visible to cosmological distances, up to a few Gpc~\citep{aba+10}.
This would provide a unique way to test General Relativity in the dynamical strong-field
regime (so far unverified by direct observations), and shed some light on  astrophysically interesting 
 properties of  the detected sources like their masses, spins and birth rate~\citep[e.g.,][]{abramovic+92}.

At present, astrophysical predictions for compact binary coalescence rates remain highly uncertain~\citep{aba+10}. 
Rate estimates mostly  relay on population  synthesis models which are based on a number of assumptions
and  poorly constrained model parameters~\citep[e.g.,][]{narayan+91,pz+yungelson98,belczynski+2002, kalogera+07}. 
These  models often predict  that 
detections  of GW sources from the galactic field will be dominated by NS binary inspirals
while  BH binaries  will only  be a small contributor to the total rate~\citep[about 1 in 10 detections;][]{sha+07}.
This conclusion  is based on results of stellar evolution calculations  which 
indicate that  formation of close BH pairs through evolution of massive stars in isolated binary
systems is a rare phenomenon~\citep{belczynski+2007}. 

Although BH binaries might be rare in the Galactic field, it is possible that they 
efficiently form via dynamical interactions in dense stellar environments~\citep{sigurdsson+phinney} 
like for example  galactic nuclei~\citep{miller+lauburg09}
and globular  clusters~(GCs)~\citep{oleary+06,DBGS10,DBGS11,BBK10}. In these dense stellar systems, otherwise
rare dynamical processes can take place and affect the dynamical evolution of
the BH population. Due to dynamical friction BHs  tend to segregate~\citep{Spitzer}
toward the core of the cluster, or galaxy, where  dynamical 
interactions with stars and other BHs  will lead to the  formation, hardening 
and (in some cases) ejection of BH binaries~\citep{Heggie}
before they  can efficiently inspiral and coalesce in response to gravitational radiation. 
The merger rate  for these binaries, although  uncertain, 
could be as large as a few thousand events a year for  aLIGO and
 be comparable or  even greater than the event rate for NS binary 
mergers~\citep[e.g.,][]{PortegiesZwart+McMillan00,sadowski+08,DBGS11}.

Since stellar BH binaries  are not directly detected,
detailed modeling of their formation processes and dynamical evolution
is crucial for making  predictions about  their characteristic GW-form and in turn 
for enabling  efficient detection  with future GW observatories.
It was first noted by \citet{mil+02e} that binary-binary encounters
play a decisive role in initiating  favorable conditions for BH binary  coalescence  in  GCs.
A large fraction, roughly  $20-50\%$, of these encounters leave behind a stable 
hierarchical (BH) triple system~\citep[e.g.,][]{sigurdsson+hernquist93,kulkarni+93}.

If the orbital plane of the inner binary is strongly tilted 
with respect to the orbital plane of the outer object the secular Kozai mechanism can lead 
the inner binary orbit to very high eccentricities~\citep{1962K,1962L}. In turn, 
given that the timescale associated with GW radiation is a strong function of eccentricity~\citep{pet64},
a combination of Kozai cycles and GW energy loss can enormously accelerate the merger of a compact binary~\citep[e.g.,][]{bla+02}.
A quite convincing demonstration of the importance of the Kozai mechanism  in determining the conditions for
 BH binary mergers in stellar clusters  was recently provided by \citet{aarseth12} who made use of
a hybrid $N-$body code incorporating  the algorithmic regularization method of \citet{hellstrom+mikkola10} and
post-Newtonian forces to evolve GC models containing BHs and NSs. 
Aarseth finds that the onset of conditions for GW radiation is usually initiated by the Kozai mechanism.
 
A number of papers have considered the Kozai resonance as a mechanism for  accelerating
compact object mergers in triple systems~\citep[e.g.,][]{bla+02,mil+02e,wen03,tho+11,prodan+13}.
These authors dealt with the problem by averaging over  orbital phases, i.e., averaging
the equations of motion over the short timescales associated with the 
unperturbed Keplerian motions of both  inner and outer orbit. 
The result is a set of first-order differential equations  that describe the long-term (secular)
evolution of the remaining orbital elements due to the perturbing forces~ \citep{david-book}. 

In this paper we perform direct integrations of the post-Newtonian  equations of motion describing  triple systems.
We do this by 
 using the AR-CHAIN code which is able to trace
the motion of tight binaries with arbitrary mass ratio for long periods of time  with
extremely high precision~\citep{MM:06,MM:08}. The code combines the use of the
chain regularization method of~\citet{mikkola+aarseth93} and the time transformed leapfrog scheme to avoid singularities.
This  allows for an essentially exact  treatment of the dynamics
 avoiding the approximations that are made when using the orbit average treatment, at expense, of course, of computational efficiency.
The employment  of  an $N-$body integrator  is motivated by a recent study of
 \citet{AP12} who have shown, in the context of BH binary mergers in galactic nuclei,  
that the orbit averaged approximation  breaks down for highly inclined but stable configurations if the binary 
comes closer to the outer perturber than a certain  distance~(see  \S~4.1 of Antonini \& Perets~2012, or Katz~\&~Dong~2013 and
Seto~2013 for an application of this in the context of white dwarf-white dwarf and NS-NS collisions respectively).
 
 In this work we investigate the dynamics  of coalescing  BH binaries driven by the Kozai mechanism and
 focus  on their eccentricity distribution when the GW signal reaches  frequencies that will be detectable by 
the next generation of ground-based laser interferometers.  Similar analysis  were previously conducted by 
\citet{mil+02e}, \citet{wen03} and \citet{oleary+06}. These previous papers employed the standard 
orbit average  approximation, which we demonstrate to be inaccurate in describing the dynamics of such systems even when 
including octupole order terms in the evolution. 
Our $N-$body experiments reveal  that  inspiraling BH binaries  often reach the gravitational-radiation-dominated regime while on orbits that 
are still very eccentric;  eccentricities can be large enough that detecting the GW signal from these sources 
will require the use of eccentric templates for data analysis and dedicated search strategies~\citep{east+13}.

In \S~2 we review the concepts behind the classic Kozai secular theory for the evolution of hierarchical triples 
noting  the simplifying approximations that are made in this treatment. In  \S~3 we show that the standard  orbit average approximation
breaks down if the inner binary is orbited by an outer perturber with a moderate periapsis, and determine an approximate condition for this to occur. In \S~4
we make use of direct numerical three-body integrations  to study the dynamics of hierarchical  BH triples in
 GCs and focus on the evolution of the inner BH binary  into a short period orbit and the way in which it may become an observable GW source
 for the advanced GW detectors. We discuss and summarize  our results in \S5.

\section{Orbit averaged equations}
We consider stellar BH binaries with components of masses $m_{0}$
and $m_{1}$ orbiting a third  body of mass $m_2$.
We denote the eccentricities of the inner and outer orbits, respectively,
as $e_{1}$ and $e_{2}$, and semi-major axes $a_{1}$ and $a_{2}$.
We  define $\omega_{1}$ as the argument of periapsis of the inner
binary relative to the line of the descending node, $\omega_2$ as the argument of periapsis 
 of the outer orbit, and $I$ as the
mutual orbital inclination of the inner orbit with respect to the outer
orbit.

If the distance of the outer BH is much larger than the semi-major axis of the inner binary
 the dynamics of the entire system can be described as the interaction 
between  an inner binary of point masses $m_{0}$ and $m_{1}$ and an external
binary of masses $M_{b}=m_{0}+m_{1}$ and $m_{2}$. 
We  define the angular momenta $L_{1}$ and $L_{2}$ of the inner and
outer binary and total angular momentum ${\mathbf H}={\mathbf L}_{1}+{\mathbf L}_{2}$:
\begin{equation}
L_{1}=m_{0}m_{1}\left[\frac{Ga_{1}\left(1-e_{1}^{2}\right)}{M_{b}}\right]^{1/2}~,\end{equation}
and \begin{equation}
L_{2}=M_{b}~m_{2}\left[\frac{Ga_{2}\left(1-e_{2}^{2}\right)}{M_{b}+m_{2}}\right]^{1/2}~,\end{equation}
with $G$ the gravitational constant.
We  define the dimensionless angular momenta
$\ell_1=L_1/L_{1,c}$ and $\ell_2=L_2/L_{2,c}$ with $L_{i,c}=L_i/\sqrt{1-e_i^2}$
the angular momentum of a circular orbit with the same $a_i$.

When  changes in the orbital properties of a
three-body system occur on a timescale longer than
both the inner binary and the outer tertiary orbital periods, it is convenient
to average the motion over both these periods.
The resulting double average Hamiltonian of the system is $\mathcal{H}=kW$
with $k=3Gm_{0}m_{1}m_{2}a_{1}^2/8M_{1}a_{2}(1-e_{2}^2)^{1/2}$
and \citep{mil+02e} \begin{eqnarray}\label{Ham}
W(\omega_{1},e_{1}) & = & -2(1-e_{1}^2)+(1-e_{1}^2){\rm cos^2}I \\
&  & +5e_{1}{\rm sin}2\omega_{1}({\rm cos^2}I-1)+\frac{\chi}{\sqrt{1-e_1^2}}~,\nonumber \end{eqnarray}
with 
 \begin{eqnarray}\label{pn}
\chi=\frac{8M_b}{m_2} \left[\frac{a_2(1-e_2^2)}{a_1}\right]^3\frac{GM_b}{a_1c^2}
\end{eqnarray}
and $c$ the speed of light.

The quadrupole-level secular perturbation equations 
can be easily derived from the conserved Hamiltonian~(\ref{Ham}).
The resulting evolution equation of the inner binary orbital elements, including 
Schwarzschild precession~(SP) and quadrupole gravitational wave radiation terms, are \citep{for+00,pet64}:
 
\begin{eqnarray}
\frac{de_{1}}{dt} & = & \frac{30K}{L_{1,c}}e_{1}\sqrt{1-e_{1}^{2}}\left(1-{\rm cos}^{2}~I\right){\rm sin}~2\omega_{1}\nonumber \\
&  & -\frac{304G^{3}m_{0}m_{1}M_{b}e_{1}}{15c^{5}a_{1}^{4}\left(1-e_{1}^{2}\right)^{5/2}}
\left(1+\frac{121}{304}e_{1}^{2}\right)~,\label{eq:eccen}\end{eqnarray}
\begin{eqnarray}
\frac{d\omega_{1}}{dt} & = & \frac{6K}{L_{1,c}}\Big(\frac{1}{\sqrt{1-e_1^2}}\Big[4{\rm cos}^{2}~I+\nonumber \\ 
&& (5{\rm cos}~2\omega_{1}-1)
(1-e_{1}^{2}-{\rm cos}^{2}I)\Big]\label{eq:g1}\\
&  & +\frac{L_{1,c}{\rm cos}~I}{L_{2}}\left[2+e_{1}^{2}(3-5{\rm cos}~2\omega_{1})\right]+
\frac{\chi}{1-e_1^2}\Big)~,\nonumber \end{eqnarray}
\begin{equation}
\frac{da_{1}}{dt}=\frac{-64G^{3}m_{0}m_{1}M_{b}}{5c^{5}a_{1}^{3}
(1-e_{1}^{2})^{7/2}}\left(1+\frac{73}{24}e_{1}^{2}+\frac{37}{96}e_{1}^{4}\right)~,\label{eqm1}\end{equation}
\begin{eqnarray}
\frac{dH}{dt} & = & -\frac{32G^{3}m_{0}^{2}m_{1}^{2}}{5c^{5}a_{1}^{3}(1-e_{1}^{2})^{2}}\left[\frac{GM_{b}}{a_{1}}\right]^{1/2}\label{eqm4}\\
&  & \left(1+\frac{7}{8}e_{1}^{2}\right)\frac{L_{1}+L_{2}{\rm cos}~I}{H}~,~~~~~~~~\nonumber \end{eqnarray}
where \begin{equation}
K=\frac{Gm_{0}m_{1}m_{2}}{16M_{b}a_{2}(1-e_{2}^{2})^{3/2}}\left(\frac{a_{1}}{a_{2}}\right)^{2}~\end{equation}
and \begin{equation}
{\rm cos}~I=\frac{H^{2}-L_{1}^{2}-L_{2}^{2}}{2L_{1}L_{2}}~.\label{alfa}\end{equation}

Equations~(5)-(8) can be considered as describing the interaction
between two weighted wires instead of point masses in orbits.
Generally, these equations are  applied to the dynamics of triple
systems that satisfy the stability criterion~\citep[e.g.,][]{bla+02,wen03,FAB,Hamers+13}: 
 \begin{eqnarray}\label{stab}
\frac{a_2}{a_1} &>& \frac{3.3}{1-e_2}\left[ \frac{2}{3}\left(1+\frac{m_2}{M_b}\right) 
\frac{1+e_2}{\left(1-e_2\right)^{1/2}}\right]^{2/5}\\
&&\times(1-0.3I/\pi)~, \nonumber
\end{eqnarray}
which was derived in ~\citet{MH01} by means of Newtonian direct $N-$body simulations.
Systems that satisfy this  criterion are stable hierarchical systems, meaning 
that the semi-major axis of the inner binary is constant on a secular timescale, in contrast to
unstable systems that experience chaotic energy exchange which inevitably  leads 
to the escape of one body over a short timescale. 

The binary starts from initial eccentricity $e_{1}(0)$, semi-major axis~$a_1(0)$, argument of periapsis $\omega_{1}(0)$
and mutual inclination $I(0)$ and evolves through interaction with the third body to a maximum eccentricity
$e_{{\rm max}}$ and critical $\omega_{{\rm crit}}$ and $I_{{\rm crit}}$.
For initially small inclinations, $60{\rm cos^2 I(0)} \gg \chi^2$, relativistic terms 
can be ignored and one finds $e_{{\rm max}}=\left[1-(5/3){\rm cos^2 I(0) } \right]^{1/2}$~\citep{Innanen+97}.
High initial inclinations will instead result  in
large values of $e_{{\rm max}}$ for which relativistic terms become relevant,
for instance by limiting the maximum eccentricity attainable by the inner binary orbit. 
 The influence of  relativistic processes on the 
dynamical evolution of the BH triple are briefly discussed in what follows.

\subsection{The role of relativistic effects}
If the inner binary precesses ``quickly", i.e., in a timescale short compared with 
the time for changes in  $\bf{L_1}$, 
the resulting, averaged equations of motion
 will no longer contain $\omega_1$, and so they 
will conserve  the momentum conjugate to $\omega_1$, which is  $L_1$. 
This simple argument indicates that any mechanism inducing rapid 
apsidal precession, e.g., SP for an eccentric orbit,
will suppress the Kozai resonance. 
However,  if the relativistic precession timescales are 
comparable or lower than the secular Newtonian  timescales, then around some critical value of $a_2$ 
it is   possible that the maximum  achievable eccentricity is increased with respect to the 
Newtonian case \citep{ford+2000,naoz+13}.
 
For highly eccentric orbits, the SP (whose associated rate diverges as $(1-e_1)^{-1}$)
sets a lower bound, $\ell_{\rm SP}$, to the angular momentum attainable by the inner binary orbit.
The minimum angular momentum allowed can then be  obtained
from the quadrupole-order equations of motion~(5)-(8) by setting $de_1/dt=0$  
and from conservation of $W(\omega_{1},e_{1})$~\citep[e.g.,][]{mil+02e}.
If we ignore energy loss due to GW emission,
 in the restricted three body problem (i.e., $m_0\gg m_1 \gg m_2$),
 one finds~\citep{mil+02e,wen03}: 
\begin{eqnarray}\label{sp_om}
\ell_{\rm SP}\approx \sqrt{1-\left(\frac{\chi}{9}\right)^2}~,
\end{eqnarray}
where  $e_1\approx 1$, high initial inclination~($I(0)\approx \pi/2$),
and weak precession ($\chi\ll 3$) have been assumed.

The importance of SP to the evolution of the inner binary orbit
depends on the role of GW emission which might
dominate  the evolution before 
$\ell_{\rm SP}$ is reached.
From Equation~(5), by setting $de/dt=0$ and neglecting angular terms, 
we find that the critical angular momentum below which 
GW energy loss dominates the evolution is approximately:
\begin{eqnarray}\label{gw_om}
\ell_{\rm GW}\approx \left( \frac{G^{3}m_{0}m_{1}M_{b}}{c^{5}a_{1}^{4}}\frac{L_{1,c}}{K}\right)^{1/6}~.
\end{eqnarray}
Transition between Kozai dynamics and GW driven inspiral occurs at
\begin{equation}\label{eq:GW-crossover}
 \ell_1=\ell_{\rm GW}.
\end{equation}
For $\ell_{GW}< \ell_1$, the inner binary ``decouples'' from the third body, and inspirals 
approximately  as an isolated system.
If $\ell_{\rm GW}>\ell_{\rm SP}$, GW radiation reaction will dominate the evolution within one Kozai cycle and the SP will be unimportant.
In the (quadrupole-level) orbit average approximation this is a necessary condition for 
the binary to maintain  a finite eccentricity as its GW frequency evolves 
through the aLIGO band.

Given the rapid orbital circularization due to GW emission
we would expect  $e_1$ to be typically  small as the binary
GW frequency passes through  the high frequency 
band of ground based interferometers.
On the other hand, it is possible that the binary GW frequency crosses
 the lower end of the aLIGO frequency 
band,~$f_{\rm GW}^L=10~{\rm Hz}$, before 
GW radiation dominates  the evolution, which will allow the eccentricity to be 
  large~($\sim e_{\rm max}$) at such high frequencies~\footnote{http://www.ligo.caltech.edu/advLIGO/scripts/summary.shtml}.
Such situation occurs only  if $\ell(f_{\rm GW}^L)>\ell_{\rm GW}$, 
with $\ell(f_{\rm GW}^{L})$  the binary angular momentum
corresponding to the detector characteristic frequency and initial binary
semi-major axis $a_1(0)$. 

Eccentric binaries emit a GW signal with a broad spectrum of frequencies;
the peak gravitational wave frequency  corresponding to
the harmonic which leads to the maximal emission of GW radiation
can be approximated as~\citep{wen03} \begin{equation}
f_{{\rm GW}}=\frac{\sqrt{GM_{b}}}{\pi}\frac{~~(1+e_{1})^{1.1954}}{\left[a_{1}(1-e_{1}^{2})\right]^{1.5}}~.\end{equation}
Under the quadrupole approximation the condition for the system to be able to  pass through the detector frequency band
before GW radiation dominates is
\begin{eqnarray}\label{hi-ecc}
\frac{a_2 \ell_2}{a_1} &<& \left(\frac{2}{5}\frac{c^5}{(\pi f_{\rm GW}^L)^2}\frac{m_2M_b}{m_0m_1}\right)^{1/3}
\frac{a_1^{-1/6}}{\left(GM_b\right)^{1/2}} \\
&=&3.3\left(\frac{1}{2} \frac{m_2M_b}{m_0m_1}\right)^{1/3} 
\left(\frac{f_{\rm GW}^L}{10{\rm Hz}}\right)^{-2/3} \nonumber \\
&&\times \left( \frac{M_b}{20M_\odot}  \right)^{-1/2}\left(\frac{a_1}{1\rm AU}\right)^{-1/6}~.
\nonumber
\end{eqnarray}
By additionally requiring  the triple system to be stable,
Equations~(\ref{stab}) and  (\ref{hi-ecc}) imply that
the range of initial conditions that would allow a  BH binary to enter the aLIGO frequency band
while on a high eccentric orbit is quite small. For instance, 
for the compact-object masses considered here,
setting $a_1= 0.3~$AU and $e_2\sim 0$, highly eccentric GW sources can only
be produced  for  $2.8\lesssim a_2/a_1\lesssim4$.
However, as we demonstrate in the next section, 
such a result is an artifact  of  the double-average approximation which is shown to break 
down for  $a_2(1-e_2)/a_1\lesssim 10$, i.e., well above the 
stability boundary implied by Equation~(\ref{stab}). In fact, we find that 
in this region of parameter space, the system  evolves through a complex dynamical evolution which 
can lead to arbitrarily large eccentricities  before the fast orbital circularization due to GW loss begins.
Our calculations predict that at least a few percent of  BH binaries
will have extremely high eccentricities while entering  the aLIGO band.

\section{direct numerical integrations}
\subsection{Numerical method}
We tested the applicability of the orbit-averaged approach  for the 
study of GW sources  by comparing the results of the orbit-averaged equations 
of motion (in both their quadrupole and octupole level 
form\footnote{We integrated the equations of motion of \citet{bla+02} by using
 a 7/8 order Runge-Kutta algorithm with a variable time-step \citep{F:68}
 in order to keep the relative error
per step in energy, in the absence of GW energy loss,
 less than $10^{-8}$. 
 When  GW  radiation was included,  we checked the integration accuracy
 through the quantity  $E+E_{\rm GW}$  with $E$ the energy per unit mass and 
 $E_{\rm GW}$ the work done by GW radiation along the trajectory.
The accuracy in this case was  
of the same order of that found in integrations without dissipative terms.}
), with those from numerical three-body integrations
having the same initial conditions (and arbitrary orbital phases).
The triple dynamics was followed until the inner binary GW signal reached the lower end of the aLIGO frequency 
band, which we identify here with the moment at which the binary peak GW frequency reaches the $10~$Hz frequency.

We used the high accuracy $N-$body integrator 
AR-CHAIN~\citep{MM:08},   which includes post-Newtonian~(PN) non-dissipative 1PN, 2PN
and dissipative 2.5PN  corrections to all pair-forces. The code employs an
algorithmically regularized chain structure and the time-transformed
leapfrog scheme to accurately trace the motion of tight binaries with
arbitrarily large mass ratios. This permits an essentially exact (at the PN level) treatment of the interplay 
between Newtonian and relativistic  perturbations to the motion, avoiding the approximations that are
made  when using the orbit average equations of motion. We refer the reader to  \citet{MM:06} and \citet{MM:08} for 
a more detailed description of   AR-CHAIN. 

Since following the inspiral of the binary all the way down to the aLIGO band 
is not practical with AR-CHAIN, we stopped  integrating the triple  once   the binary orbital separation had
shrunk (because of GW energy loss) to about $10^{-2}\times a_1(0)$.\footnote{The semi-major axis and eccentricity were calculated  from 
the particle positions and velocities by adopting the relative
radial expressions given in Equations~($3.6~a$) and ($3.6~b$) of \citet{dam+85}.}
At this point  the binary dynamics is dominated by GW radiation and the influence of the third body 
can be safely  neglected, so we continued evolving the binary eccentricity and semi-major axis  by using the leading order 
orbit average analytical formulas of \citet{pet64}. 
This  allowed for a more efficient computation of the  dynamical evolution of the system.

 \subsection{Breakdown of the orbit average approximation}\label{BOV}
We begin by  studying  the eccentricity of  the inner binary at the time its GW signal first enters the aLIGO frequency band, i.e.
$f_{\rm GW}=10~{\rm Hz}$,  for a set of illustrative systems that are close to the stability criterion given by Equation~(\ref{stab})
and for which the initial mutual inclination is large. 
It is from these marginally hierarchical configurations that we  expect the larger number of eccentric sources for 
aLIGO, as well as the larger discrepancy between the results of $N-$body integrations 
and the predictions of  the orbit average equations of motion~\citep{AP12}.

\begin{figure}
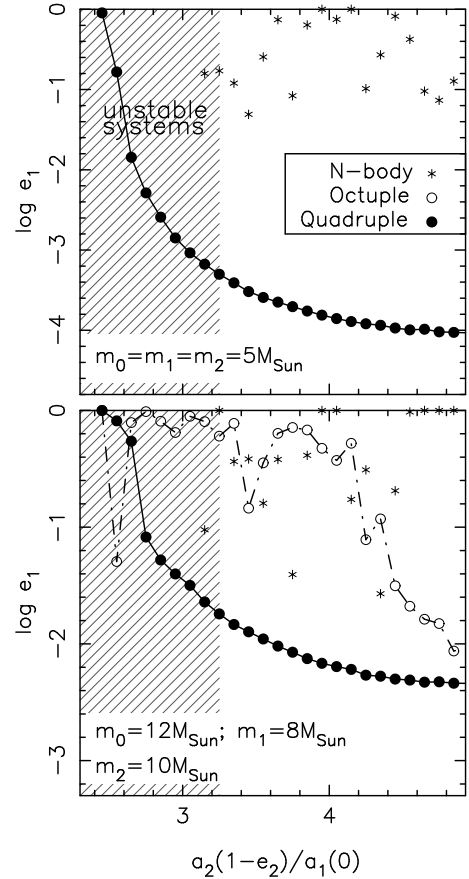
\centering
\begin{tabular}{c}
\includegraphics[clip,width=0.7\columnwidth]{fig1b.ps}
\\\includegraphics[clip,width=0.7\columnwidth]{fig1a.ps}
\end{tabular}
\caption{\label{fig1}  The eccentricity at the moment the inner BH binary 
enters  the sensitivity window of planned ground-based interferometers 
as a function of the periapsis distance of the outer body, $a_2(1-e_2)$. 
Inside the hatched regions triple systems are  unstable
 according  to Equation~(\ref{stab}); our $N-$body experiments verify the accuracy of this stability criterion. 
  In these integrations we fixed the initial inner and outer orbit semi-major axes, while
 we  distributed  the external orbit eccentricity uniformly within  the
range $e_2=(0.51,~0.01)$, corresponding to the interval of periapsis distances $a_2(1-e_2)/a_1(0)=(2.45,~4.95)$AU. 
We randomly  chose  the initial true anomalies, set $\omega_1=\omega_2=0$ and $I=99^{\circ}$.
In the upper panel,  we set $a_1=2~$AU, $a_2=5\times a_1$.
Since the  inner binary components in  these systems   have equal   masses
the octupole order terms in the secular equations of motion go to zero. The corresponding results from the secular code 
provide a  poor match to the results of the $3$-body integrations, which systematically predict
higher residual eccentricities.
In the lower panel we set $a_1=0.2~$AU and $a_2=5\times a_1$.
In these integrations $m_0\neq m_1~$, thus leading to a non-zero contribution of the octupole order terms. 
The results of the secular theory in this case
are in better agreement with the results of the $3$-body integrations. 
However,  while in the $N-$body runs a large fraction of
binaries  are  exremely  eccentric~($e_1\gtrsim0.9$) when the first enter the aLIGO band, the residual eccentricity of such binaries is substantially smaller~($e_1\sim0.1$)  
when the same initial conditions are evolved using the orbit average equations of motion (even when  high order terms are included). 
Evidently, the standard secular theory fails in  describing   the dynamical  evolution  of these systems.  }
\end{figure}

Two sets of initial conditions are explored in Figure~\ref{fig1}.
 In the upper panel we set
$m_0=m_1=m_2=5~M_\odot$. In the secular theory 
the octupole order terms go to zero for $m_0=m_1$, and we are therefore 
left with the contribution from the  quadrupole order terms only, i.e., Equations~(5)-(8).  

Systems that are unstable according to Equation~(\ref{stab})
(hatched region in  Figure~\ref{fig1}) experience the ejection of 
one component before the inner binary can complete one Kozai cycle and possibly merge.
 An example of such systems is the first point on the left in the upper panel of Figure~\ref{fig1}, which  is the same system shown in Figure~4
of \citet{wen03}. As Wen noted,
such a system would enter the aLIGO band with a large eccentricity, $e_1\approx0.9$. This occurs
because GW radiation dominates within one Kozai cycle and  SP  is negligible. 
In reality, such initial conditions correspond to 
 a highly unstable system and no merger occurs when the system is evolved in time by using the direct integrator. 
In fact, according to Equation~(\ref{hi-ecc}), the quadrupole level secular equations of motion
should, and our numerical integrations show that they do, result
in a very small residual eccentricity for all stable configurations.
In contrast 
the results of the direct $3$-body calculations give large residual eccentricities; 
 a few binaries  enter the aLIGO band with extremely  large eccentricities, $e_1\gtrsim 0.9$.   
  
We investigate the importance of  the octupole level terms in the lower panel of Figure~\ref{fig1}
where we set $m_0=8\msun$, $m_1=12\msun$ and $m_2=10\msun$. 
In this case there is a non zero
 octupole order perturbation which changes the rate of precession slightly, and causes variation in the outer
perturber orbital eccentricity.

When expanded to the quadrupole order only, 
the averaged perturbing potential is axisymmetric~\citep[even for an eccentric external orbit,~e.g.,][]{KDM11}.
A direct consequence of this is that the magnitude of the inner orbit's angular momentum 
has a well defined lower bound which under the test particle approximation is simply the Kozai constant, i.e. 
$\ell_1\leq \ell_z=\sqrt{1-e_1^2}\cos(I)$, with  $z$   chosen along the direction of $\mathbf{L}_2$.
At the octupole order the Hamiltonian  describes a dynamical
system of two degrees of freedom.   
Motion in this case differs  in three important ways from motion in the 
(axisymmetric) quadrupole order potential.
First,  the lack of rotational symmetry implies that no component of the binary angular momentum is conserved.
As a consequence of this the eccentricity of the inner orbit  can reach 
much higher values near the maximum of a Kozai cycle~\citep{for+00}.
Second, while at the quadrupole level orbits  maintain their sense of
rotation, in the octupole potential  excursions to very high
eccentricities can be  accompanied by a ``flip'' of the orbit with respect
to the direction of the total angular momentum~\citep{naoz+11a,lithwick+naoz11,naoz+11b}.
Third,  motion  in the octupole 
potential can be stochastic, 
although only a  restricted portion of the phase space near 
the libration/circulation separatrix 
might be  expected to be  chaotic~\citep{holman+97}.

\begin{figure}\centering
\includegraphics[clip,width=.9\columnwidth]{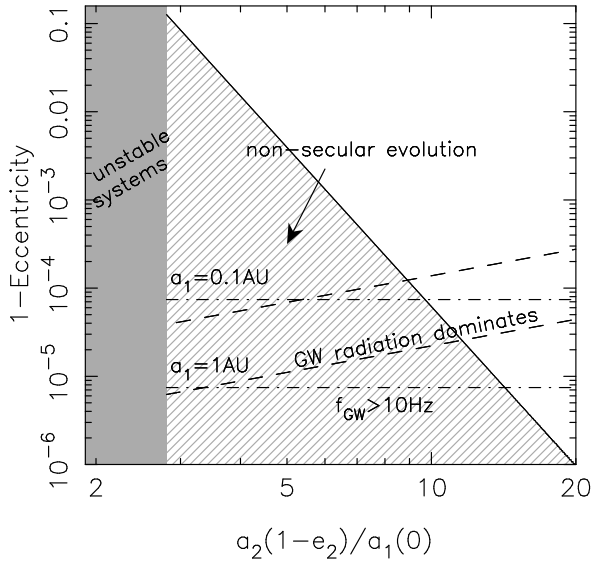}
\caption{\label{fig2} 
Illustrating the region of parameter space within which highly eccentric 
GW sources can be produced.
The solid line gives the value of the critical eccentricity 
defined in Equation~(\ref{sp_om2}) as a function of the periapsis distance of the outer BH orbit
expressed in units of $a_1$ and assuming $M_b=2m_2$. Below this line
a binary  can reach $e_1\sim1$ in a fraction of its orbital period. 
The gray solid area corresponds to unstable configurations  according to Equation~(\ref{stab}) 
where we set $e_2=0$ and $I=90^\circ$. 
 Dashed lines enclose the 
  region of parameter space within which GW radiation reaction 
dominates  the evolution. 
These lines were
computed using the approximate Equation~(\ref{gw_om}) with
$e_2=0$, $m_0=m_1=m_2=10 \msun$ and  $a_1=0.1$ and $1$~AU. 
 Within the dot-dashed lines the binary peak  GW frequency 
becomes  larger than $10~$Hz.
 Highly eccentric GW sources are produced by
 stable-moderately hierarchical  triples with $3\lesssim a_2(1-e_2)/a_1 \lesssim10$. These
are the only systems that can penetrate the region  
delimited by  the hatched area in the plot  before attaining  $\ell_{\rm GW}$.  }\end{figure}

\begin{figure}\centering
\includegraphics[clip,width=0.7\columnwidth]{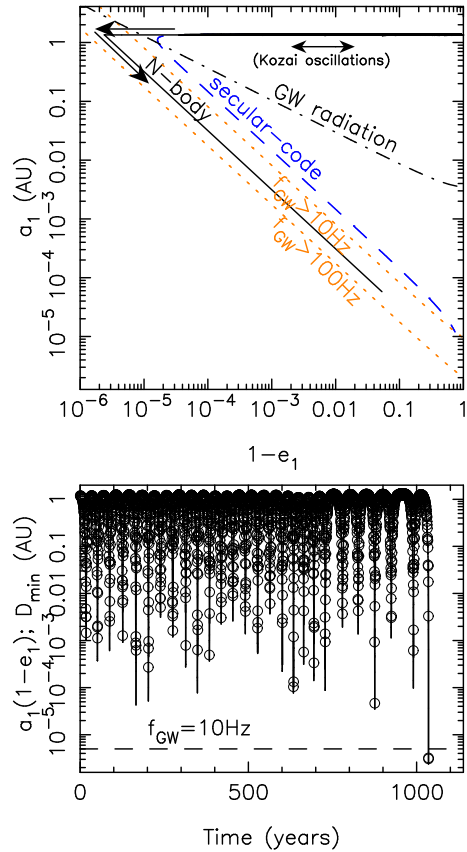}
\caption{\label{fig3}   Kozai-driven
evolution of a BH binary  in the semi-major axis-eccentricity plane.
Initial conditions are: $m_0=8~M_\odot$,~$m_1=12~M_\odot$, $m_2=10~M_\odot$; $I=97^{\circ}$;
$a_1=1.36~$AU, $a_2=6.26~$AU; $e_1=0.01$, $e_2=0.2$; $\omega_1=\omega_2=0$.
Short-dashed lines correspond to the indicated reference peak gravitational 
wave frequencies of $10$ and $100~$Hz. The 
dot-dashed  line in the upper panel gives the approximate condition 
for GW energy loss to dominate  the evolution, Equation~(\ref{eq:GW-crossover}).
To the left and below this line the BH binary eccentricity and semi-major axis 
rapidly  decrease when the system is evolved using the double average (octupole-level)
equations of motion~(blue-dashed line). 
Arrows  schematically indicate the direction in which the binary evolution proceeds.
The black-solid line
displays the evolution of the BH binary obtained by using more
accurate  $3$-body integrations.
After multiple periodic oscillations in the eccentricity have occurred~(see lower panel),
the binary reaches very high eccentricities and chirps in the aLIGO frequency
band. The orbital inspiral of the BH binary  in this case starts 
and takes place entirely within the aLIGO band.
The lower panel shows time evolution of  the actual distance of  closest approach, 
$D_{\rm min}$, between the  inner binary components~(open circles) 
and the periapsis distance, $a_1(1-e_1)$,
calculated from the current value of the osculating orbital elements~(solid line).
Near the end of the simulation the periapsis   distance 
between the BHs changes such that the GW signal enters the aLIGO band in 
one orbital period before the rapid orbital circularization  due to GW can start.
This large angular momentum change, achieved in one orbital period,
 allows the  BHs to attain a close approach and the GW signal  to enter the aLIGO  band 
at high eccentricity. The peak GW frequency at the moment the insipral phase begins
is significantly larger than $10~$Hz, about $45~$Hz in this case.}
\end{figure}

Figure~\ref{fig1} shows that when expanded at  the octupole level the secular theory
leads to higher residual eccentricities and also appears to be in better agreement
with the  results of AR-CHAIN. This suggests that 
the excitation of the inner binary eccentricity seen also in the direct integrations
depends on the importance of the Newtonian octupole~(or even higher) 
order terms to the evolution which break the axisymmetric nature of the potential and
lead to large eccentricities without the need of high inclination tuning.

From Figure~\ref{fig1} it is evident that the results of the direct $3$-body 
integrations differ from the outcome of the secular code at least in one important way:
 while  in the $N-$body runs for about $30\%$ of the binaries   the eccentricity remains 
 high~($e_1\gtrsim0.9$) when they first enter the aLIGO frequency band, the residual eccentricity of such binaries is  small
when the same initial conditions are evolved using the orbit average equations of motion. 
Note also  that the reliability of the secular approximation 
becomes progressively worse as $e_2$ (or equivalently, the contribution
of the octupole order terms) approaches zero.

The discrepancy arises because, near a minimum in $\ell_1$,
the time
\begin{equation}\label{timeKz}
\left| \frac{1}{\ell_1} \frac{d\ell_1}{d t} \right|^{-1}\approx T_{\rm b}  \frac{1}{5\pi}\frac{M_{b}}{m_2} \left[\frac{a_2 (1-e_2)}{a_1} \right]^{3} ~\sqrt{1-e_1}
\end{equation}
 for the orbit to change from its current value of $\ell_1$ to $\ell_1 \sim 0$ becomes shorter than
the binary  orbital period, $T_{\rm b}=2\pi\sqrt{a_1^3/(GM_b)}$. In deriving Equation~(\ref{timeKz}) we have 
assumed a fixed outer perturber, and
adopted the instantaneous quadrupole order torque~\citep[e.g.,][]{KD13}  taking the relevant limit~$e_1\rightarrow1$
and assuming maximal torque at the periapsis approach.
This  condition  can then be expressed in terms of the system semi-major axes and
eccentricities as:
\begin{eqnarray}\label{sp_om2}
\sqrt{1-e_1} \lesssim 5\pi \frac{m_2}{M_b}\left[ \frac{a_1}{a_2 (1-e_2)} \right]^{3}= \\
8\times10^{-3}\left( 2 \frac{m_2}{M_b} \right)\left[10 \frac{a_1}{a_2(1-e_2)} \right]^3.~~~~ \nonumber
\end{eqnarray}
 If the binary angular momentum becomes smaller than  this critical value
 the orbit can evolve  substantially  and reach $e_1\sim 1$ on a timescale short
compared to the binary orbital period, implying that  the orbit average approximation formally breaks down.
For objects of comparable masses what  determines whether or not this condition is satisfied 
is  the value of the ratio  $a_2 (1-e_2)/a_1$  in Equation~(\ref{sp_om2}) which 
parametrizes  the size of the external orbit compared to  the inner binary orbit.
The tighter the outer orbit the larger the torque produced on the inner BH binary, which  implies that 
smaller eccentricities are sufficient   to meet the criterion given in Equation~(\ref{sp_om2}).


More specifically, what happens will
depend on whether changes of the binary angular momentum occurring in one orbital period
are of order or larger than the  angular momentum associated with the scale at which other
dynamical processes~(e.g. SP or GW radiation)  can affect the  evolution.
As an example, consider the most relevant case in
which, during the high-$e$ phase of a Kozai cycle, 
the binary angular momentum 
reaches  the critical value  defined in Equation~(\ref{sp_om2})
  before SP and GW radiation terms can become important.
Since the torque due to the outer BH is maximal  near apoapsis 
while GW emission is peaked  at periapsis, it is possible that 
  a large  change of angular momentum  attained near apoapsis  reduces  
$\ell_1<\ell_{\rm GW}$, and does this in a fraction of the orbital period before 
relativistic terms can affect the evolution. If  $\ell_{\rm GW}$ is crossed while the BHs are near apoapsis, and  
the binary attains periapsis before $\ell_{\rm GW}$ is passed over,
the value of the angular momentum at which  energy dissipation due to GW emission
becomes important can  be significantly smaller than $\ell_{\rm GW}$.
This allows the binary to have large eccentricities at large GW frequencies.
Figure~\ref{fig2} illustrates the region of parameter space inside which binaries
can undergo the non-secular evolution discussed above and attain very 
large eccentricities before GW radiation dominates  the evolution. 
These binaries reside in  highly inclined-moderately hierarchical   configurations 
with $3\lesssim a_2(1-e_2)/a_1 \lesssim10$. 

An example   is given in Figure~\ref{fig3}, which 
displays  the time evolution of semi-major axis and eccentricity 
of a BH binary merger driven by the Kozai mechanism. Toward the end of the simulation, 
 the binary achieves
a large enough  change in angular momentum
such that its peak GW frequency  crosses the aLIGO frequency band in a fraction of an orbital period. 
Encounters occurring before this time are at large enough distance that dissipative and non-dissipative 
relativistic  corrections are both  negligible. The semi-major axis of the binary
 remains approximately unchanged  during the evolution before GW radiation initiates the inspiral. 
In the case shown the inspiral of the BH binary due to GW energy loss
starts when its peak GW is about   $45~$Hz, or, equivalently, when the  periapsis distance is 
approximately $10M_b$ in geometric units, 
and therefore takes place entirely within the $10~$Hz frequency band.
In contrast, when using the double average equations of motion, 
which do not contain information about orbital phases, the maximum   eccentricity attainable
by the BH binary   is artificially 
limited by GW energy loss near $\ell_{\rm GW}$. 

The evolution of the GW signal emitted by highly eccentric binaries in the aLIGO frequency band 
was discussed before in \cite{O:09} and ~\cite{kocsis+levin12}.
The eccentric signature of these binaries
would make them distinguishable from other GW sources; their GW signal will consist  of an initial phase of repeated bursts 
for minutes to days  emitted 
during periapsis approach where the GW emission is maximized, followed by a continuous  chirp signal and an eccentric merger.
The relevant timescales that determine the GW waveform are 
the orbital time,~$\Delta t_{\rm orb}=2\pi\sqrt{a_1^3/GM_b}$, and the duration of
the periapsis passage,~$\Delta t_{\rm per}=\pi\sqrt{\left[a_1(1-e_1)\right]^3/GM_b}$. 
The characteristic duration  of the GW bursts is  $\Delta t_{\rm per}$, while $\Delta t_{\rm orb}$ is
 the time between two consecutive bursts. For
 $\Delta t_{\rm per} \ll \Delta t_{\rm orb}$, the waveform consists of a train of 
 GW bursts of characteristic duration $\Delta t_{\rm per}$ emitted 
 quasi-periodically every $\Delta t_{\rm orb}$.
 Later, when $\Delta t_{\rm per}\sim \Delta t_{\rm orb}$, 
 the signal becomes continuous in time domain.
 For the system of Figure~\ref{fig3} we find 
 $\Delta t_{\rm orb}=1.8\times 10^{5}$ and
 $\Delta t_{\rm per}=3.4\times 10^{-4}$ minutes at the instant 
 where the GW-driven inspiral  of the inner binary begins, and   
 $\Delta t_{\rm orb}=5.1\times 10^{-2}$ and $\Delta t_{\rm per}=3.1\times 10^{-4}$ minutes
when the binary eccentricity has decreased to $e_1=0.95$. 

\section{Coalescing black hole binaries in globular clusters}
Mergers of  BH binaries forming  via dynamical interactions in  GCs
might dominate   the   total BH-BH merger event rates for  
aLIGO and  are therefore of wide astrophysical interest~\citep{PortegiesZwart+McMillan00,aba+10}.
Here, we determine the type and 
properties of  BH binary mergers  in GCs induced by  a combination 
of Kozai cycles and GW energy loss.
Similar studies were previously conducted by 
\citet{mil+02e}, \citet{wen03} and \citet{oleary+06}. These  papers, however, were  limited to the traditional 
orbit average secular theory which we have shown  to be inaccurate 
in describing the evolution of such systems.
To overcome the limitations of the orbit average  approach 
 we perform in what follows  direct integrations of the PN equations of motion describing hierarchical 
 BH triples. \subsection{Initial setup}\label{IS}
 
The simulation initial conditions were generated as detailed below.
The most important quantities  that affect the maximum eccentricity 
a system can reach (and consequently  its merger time) are $a_1$ and $a_2/a_1$.
For example, the details of the mass parameter distributions 
 have a relatively little effect on the eccentricity a system can reach.
Monte Carlo  simulations  suggest a distribution of
periods for BH binaries inside globular clusters which is approximately 
flat in log space~\citep[e.g.,~Figure~6 of][]{DBGS10}.  
Accordingly, in a first set of simulations~(MOD1 hereafter) the initial BH binary  semi-major axis, $a_1$,
was drown from a distribution which is flat in $\log$ space, $f(\log p)\propto$const, which corresponds 
to $f(a)\propto 1/a$~(i.e., \"{O}pik's law). The lower and upper limits for $a_1$ were $0.2$~AU and $30$~AU 
respectively. The lower limit of $0.2$~AU corresponds roughly  to the minimum value 
of the semi-major axis distribution of BH binaries obtained via Monte Carlo simulations 
in \citet{DBGS10};
the upper limit of $30~$AU is the same adopted in \citet{wen03}. The outer orbit semi-maor axis, $a_2$, was also
log-uniformly distributed. Thus we assumed that the orbital semi-maor axis
of the third component in a given triple could be chosen from a similar  distribution
of the parent binaries.   We imposed an upper limit of $a_2=30 \times a_1$,
given that the parameter range for $a_1$ decreases with increasing $a_2/a_1$~\citep{bla+02,wen03}: 
for $a_2/a_1>30$ the binary merger time (Kozai timescale) becomes typically much longer than the collision  time with
other field stars~(Equation~[\ref{coll}] below).
The initial eccentricities, $e_1$ and $e_2$, were sampled from a 
thermal distribution, $N(e)\propto e$. 
The physics behind such $e$ and $a$ distributions is well known: it is the results of energy relaxation
in few body gravitational interactions which affect the binary/triple dynamical evolution.

We adopted a random distribution in $\omega_{1}$ and $\omega_{2}$ and random
in ${\rm cos}(I)$ with $85^{\circ} \leq I \leq 110^\circ$, as required to obtain extremely large eccentricities. 
Orbital phases were also randomly distributed.
Since  unstable triples are  likely  to dissociate before completing one  Kozai cycle~(e.g., Figure~\ref{fig1}),
we only considered  configuratios  which satisfied the stability criterion given by Equation~(\ref{stab}).

We sampled the  BH masses  from the exponential distribution
\begin{eqnarray}
N(m)={\exp\left(M_c/M_0\right)\over{M_0}} \exp\left(-m/M_0\right)
\end{eqnarray}
for $m\ge M_c$ and $N(m)=0$ otherwise.
 This choice of the mass distribution is motivated by theoretical expectations  based on the energetics 
and dynamics of supernova explosions~\citep{fryer+kalogera01}.
The values of the mass scale in the exponential  and the cutoff mass are respectively $M_0=1.2~\msun$
and $M_c=6.32~\msun$~\citep{ozel+10}. This choice of parameters  is consistent 
with constraints  from observations  of Galactic soft X-ray transients 
which indicate a significant  paucity of BHs with masses less than $\sim 5~\msun$~\citep{bailyn+98,ozel+10,farr+11}.

 Although the distributions  assumed above are quite reasonable,
 we stress  that exhaustive studies
have not been made for the results of binary-binary interactions,
and  information about whether the stable hierarchical
triples that result have $a_1$, $e_1$, $a_2$, and $e_2$ distributions that
are similar to the distributions of the parent binaries remains unclear. This might therefore represent 
a source of uncertainty  for our study. 
In order  to understand the dependence of our results on the assumed initial conditions,
we perform an additional  set of simulations~(MOD2 hereafter) where we 
adopted a uniform distribution in both $a_1$ and $a_2$, a uniform distribution in eccentricities
and set $m_0=m_1=m_2=10~\msun$.
 Other parameters  and  limits of the various distributions were the same as defined above. 
This second set of initial conditions  contain a larger number of wider configurations
with larger outer periapsis separation with respect to MOD1;
on the basis of our previous analysis~(e.g., Figure~\ref{fig2}) we would therefore expect 
the secular orbit average code to provide a better description of the BH-triple dynamics
in this case and the coalescing binaries 
to have smaller eccentricities in the aLIGO band. MOD2
corresponds to our most conservative assumptions.

 Due to the crowded stellar environment of GCs, triple systems might 
be perturbed through encounters with other stars on timescales 
that are shorter than the relevant secular timescale~(Kozai evolution).
Such encounters will alter the orbital properties of the triple significantly or even disrupt it.
To account for this we set the final
integration time in our simulations equal to the timescale,
\begin{equation}\label{coll}
T_{\rm coll}=2\times 10^5~{\rm yr} \left(\frac{10^6~{\rm pc}^{-3}}{n}\right)
 \left(\frac{\rm AU}{a_2}\right)\left(30\msun \over {M_b+m_2}\right)~,
\end{equation}
for collisions with  field stars~\citep{bt87}, where $n=10^6~{\rm pc}^{-3}$ is  the 
number density of stars in the GC~\citep[e.g.,~][]{mil+02e,wen03}.
We  first evolved the systems forward in time by using 
the octupole level orbit average equations of motion and obtained an estimate of  the binary
merger time, $T_{\rm AV}$.
We  discarded systems 
that after  a time $T_{\rm coll}$ did not produce a merger event, while
initial conditions  that  successfully lead to a merger, i.e., 
\begin{equation}\label{coll_crit}
T_{\rm AV}<T_{\rm coll},
\end{equation}
 were  realized as point-mass  particle representations and evolved  forward in time by using AR-CHAIN.
The maximum time of integration in the $3$-body runs was also set to $T_{\rm coll}$.

Strictly speaking, since the orbit averaged approach is inaccurate in determining the merger properties of the triple,
selecting systems on the basis of Equation~(\ref{coll_crit}), although computationally  
convenient, might introduce some artificial bias 
in our results.   To check on this, we run an additional set of direct simulations~(MOD1-bis hereafter) 
in which we only consider systems which satisfy the condition  $T_{\rm NB}<T_{\rm coll}$, with $T_{\rm NB}$ the merger 
time of individual systems computed via direct $3$-body integrations.
Thus, the initial conditions for the $3$-body integretions in MOD1-bis were not selected on the basis of Equation~(\ref{coll_crit}), 
rather  they were   sampled from the entire distribution corresponding to MOD1,
and directly integrated  with AR-CHAIN until  either the inner binary  merged, the triple was dissociated,
or the maximum time, $T_{\rm coll}$, was reached.
As shown in the analysis that follows,  
the correspondence between the results of MOD1 and MOD1-bis is good, assuring us 
that general conclusions of our paper are indeed not particularly  sensitive to the assumed initial conditions
and/or methodology adopted to generate the initial conditions. 
We stress, however,  that given the idealized initial conditions used in this paper and
the uncertainty in the properties of BH triple systems in GCs our results should be only considered
as a set of baselines for making predictions about the GW signal produced by such systems.
In total, we run  $3000$ $3$-body simulations.

\begin{figure}\centering
\includegraphics[clip,width=.7\columnwidth]{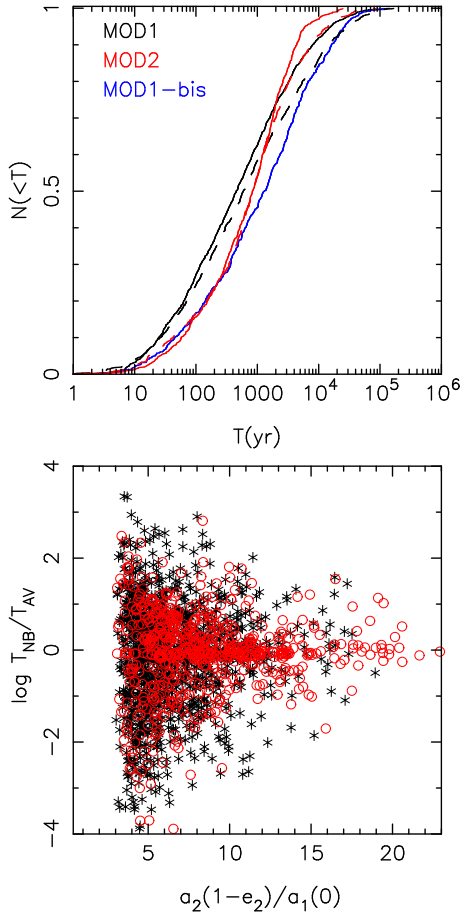}
\caption{\label{fig4} Top panel shows the cumulative distribution of merger times
obtained via direct numerical integration of the
equations of motion~(solid lines) and from the orbit average Hamiltonian model~(dashed lines).
 Lower panel gives the actual  merger time of individual systems~($T_{\rm NB}$) divided by the same timescale 
 when computed  by using  the secular orbit average code~($T_{\rm AV}$), as a function of the  periapsis separation
 of the outer BH.
}\end{figure}

\subsection{Results}
\subsubsection{Merger time distributions}
In MOD1 we found that $61\%$ of systems that merged when evolved  
using  the double average  equations of motion also merged  when  evolved  via direct $N-$body integrations.
In MOD2 this percentage increases, but only slightly, to $70\%$. 
The upper panel of Figure~\ref{fig4} compares the cumulative distribution of merger times for BH binaries calculated via  direct integrations
with  merger times obtained by using the orbit average code. The two methods generate similar merger time distributions.

The bottom panel in Figure~\ref{fig4} 
shows the merger time of individual systems, $T_{\rm NB}$, which we divided by the 
merger time computed  by using  the   orbit average model and plotted as a function of $a_2(1-e_2)/a_1$.
Most  mergers occur at $3 \lesssim a_2(1-e_2)/a_1\lesssim 10$,
a regime where the assumptions behind the secular  approach are not valid.
As a result the merger time obtained via the standard Kozai treatment can be inaccurate by
several orders of magnitude; this is especially true for the systems with the shortest periapsis separations.
 In MOD1~(MOD2) we found  that $13\%$~($10\%$) of binaries  have $T_{\rm NB}/T_{\rm AV}>10$ and $9\%$~($10\%$) have 
 $T_{\rm NB}/T_{\rm AV}<0.1$. These  results suggest that the orbit average Hamiltonian model can lead
 to misleading results and to an incorrect determination of  the  merger time of BH binary mergers in GCs
(a point we return to  in \S~\ref{conc}).

The merger time distribution 
of MOD1-bis is slightly biased towards longer merger times when compared to the
results of the orbit average code. 
On the basis of our previous analysis, this is indeed not surprising.
As described in the previous section the orbit
average approximation becomes unreliable  
if the angular momentum of the  inner binary changes on a time which is comparable or shorter than the inner
binary orbital period, $T_{\rm b}$. In the orbit average integrations 
systems that satisfy such condition
typically merge when they first enter the high eccentricty 
phase of a Kozai cycle. Thus, the typical merger timescale  of these systems
will be the time scale to reach the maximum eccentricity of a Kozai cycle, i.e.
roughly the Kozai time scale.
In the N-body integrations instead, the binary angular momentum can
reach and pass over the $\ell_1<\ell_{\rm GW}$ region 
before the binary attains periapsis. If this is the case, it might take many Kozai 
cycles before the two BHs  attain a periapsis distance such that
the GW radiation dominated phase can start.

\begin{figure*}\centering
\begin{tabular}{c}
~~~~~~~~~~~~~~~\includegraphics[width=.34\linewidth,angle=270]{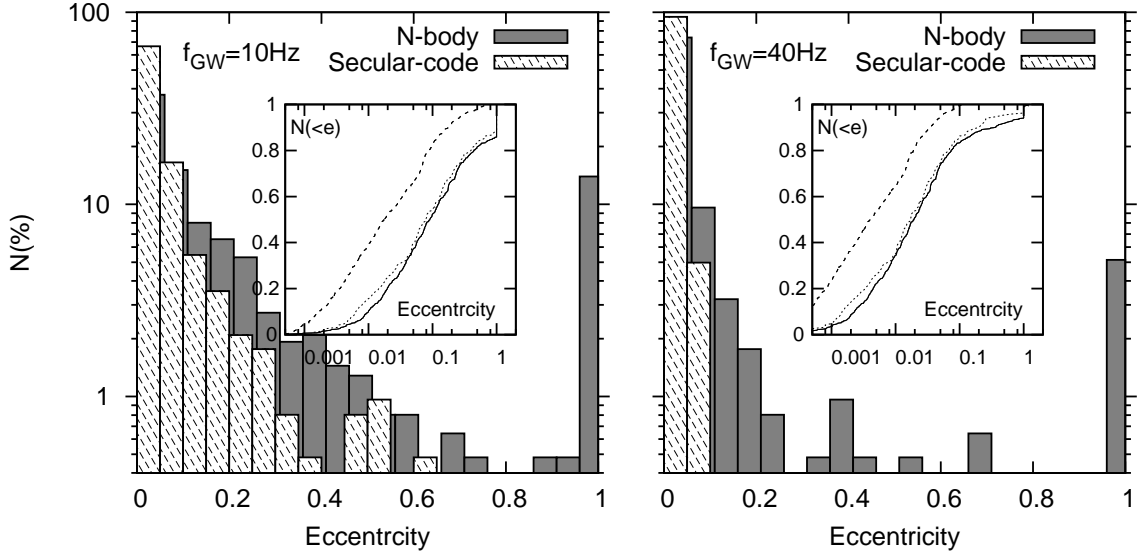}\\ \\ \\ \\\\
\end{tabular}
\caption{\label{fig5} Histograms of the eccentricity distribution in MOD1 of coalescing BH binaries driven by the Kozai mechanism in GCs
when their emitted GW signal first enters the $10$~Hz~(left panel) and the $40$~Hz~(right panel)~frequency bands. Insert panels
show the corresponding cumulative distribution of eccentricities. 
 Distributions are normalized to the total number of systems that merge
before being perturbed by encounters with other stars in the cluster. Systems were evolved using both the octupole-order orbit 
average equations of motion~(hatched histograms and dashed lines) and more accurate $3$-body direct integrations~(filled histograms and 
solid lines).
The $3$-body runs produce orbital distributions that are significantly more biased towards large eccentricities, and predict that about 
$50~\%$ of merging binaries have eccentricities larger than $0.1$ and that 
$10~\%$ of them posses extremely large eccentricities~($e_1\sim1$) when they first enter the aLIGO frequency band. The number of these highly 
eccentric GW sources is strongly underestimated when using the orbit average (secular) code.
Dotted lines in the insert panels give the cumulative distribution of eccentricities of coalescing BH binaries in MOD1-bis.
In this model the $3$-body initial conditions  were sampled from the entire distribution corresponding to MOD1,
instead of  being selected on the basis of Equation~(\ref{coll_crit}). 
}
\end{figure*}

\begin{figure*}\centering
\begin{tabular}{c}
~~~~~~~~~~~~~~~\includegraphics[width=.34\linewidth,angle=270]{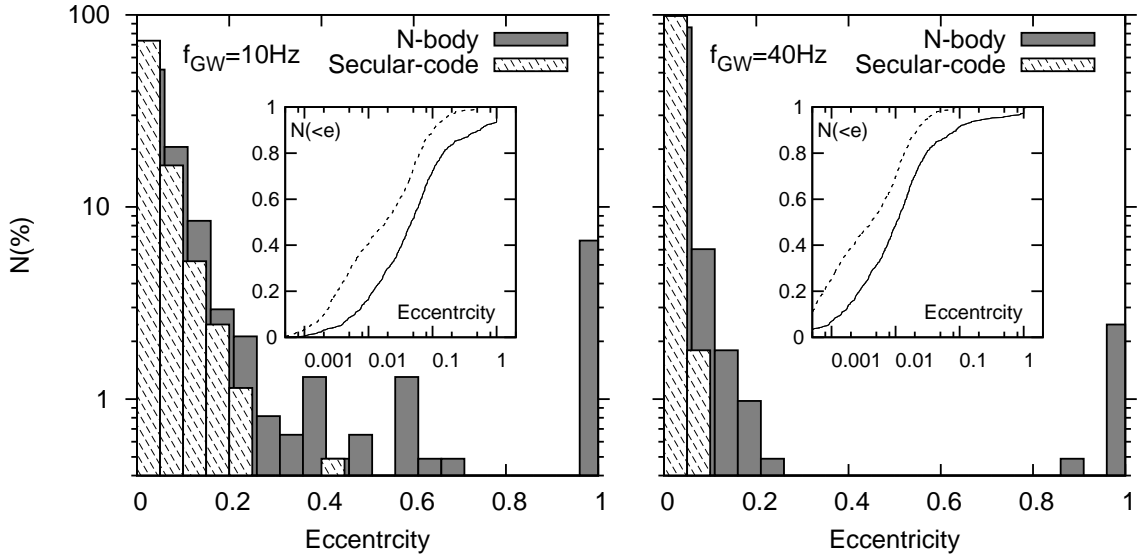}\\ \\ \\ \\
\end{tabular}
\caption{\label{fig6} Same as Figure~\ref{fig5} but  adopting an initial  uniform distribution 
in semi-major axes and eccentricities~(MOD2).}
\end{figure*}

\subsubsection{Eccentricity distributions}
The left and right panels of Figure~\ref{fig5} give the number and cumulative distributions
 of eccentricities for MOD1 at the moment the GW signal enters the $10$~Hz
and the $40$~Hz~ frequency bands respectively. The histograms are normalized to the total number of systems that manage to merge before being significantly perturbed by encounters with field stars.
MOD1 and MOD1-bis produce similar residual eccentricity distributions with
about $50\%$~($20\%$) of merging systems with eccentricities larger than $0.1$ 
 and roughly  $10\%$~($5\%$) of them with extremely high eccentricities, $1-e_1\lesssim 10^{-5}$,
at the moment  they enter the  $10~$Hz~($40~$Hz) frequency band. 
The reason for the gap in the eccentricity distributions between $e_1\sim0.6$ and $1$ resides in the fact that when a BH binary inspiral begins outside the $10~$Hz requency then it remains outside this frequency band as it circularizes and evolves  at roughly constan periapsis and peak gravitational wave frequency (e.g., dashed blue line in Fig. 3). Thus, even an initially very eccentric binary will be typically circularized by the time it  reaches the 10Hz frequency. 

Results of  integrations of MOD2,   in
which eccentricities and semi-major axes follow uniform distributions,
are displayed in Figure~\ref{fig6}.
In this case we found that roughly $30\%$~($10\%$) of merging systems have eccentricities larger than $0.1$ 
 and   $7\%$~($3\%$) of them have extremely high eccentricities, $1-e_1\lesssim 10^{-4}$,
at the moment  they enter the  $10~$Hz~($40~$Hz) frequency band. 
Due to the larger values of $a_2(1-e_2)/a_1$ adopted  in this second set of 
initial conditions  the residual eccentricity  distributions  
contain less  eccentric orbits than in Figure~\ref{fig5},   in agreement with the results of \S~\ref{BOV}.
However, the basic results of our study, i.e., a large fraction of coalescing binaries with a finite eccentricity and about $10\%$
of them with an extreme eccentricity, remain essentially unchanged. 
After this paper was submitted, \citet{antognini+13} presented a similar study on the breakdown of the secular approximation and found
a similar fraction of eccentric  compact object binary mergers.

The initial conditions  for MOD2  are somewhat more similar to what was adopted in \citet{wen03}.
Using the quadrupole order orbit average equations, Wen finds that about $2~\%$ of binaries have 
$e_1\sim 1$ at $10~$Hz.
When using the double average equations of motion, we  find,
contrary to  Wen,  that the percentage of such high eccentric sources is only $0.1\%$
of the total number of coalescing binaries.

\begin{figure}\centering
\begin{tabular}{c}
\includegraphics[width=.8\linewidth,angle=270]{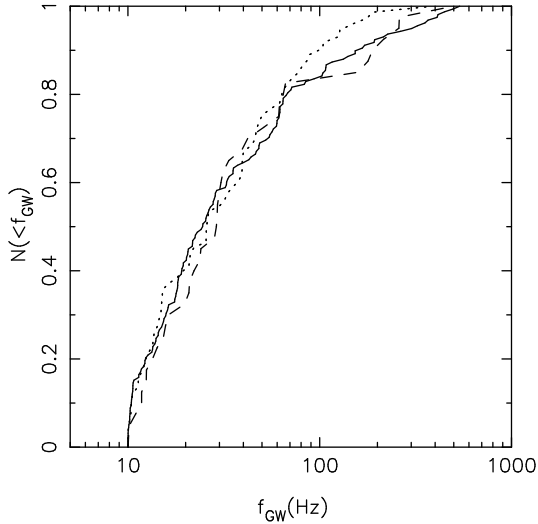}
\end{tabular}
\caption{\label{fig7} 
Dominant GW frequency at the instant where the GW-driven inspiral of inner binary begins to dominate over the Kozai-dynamics. Shown is a subset of the $3$-body configurations from  
Figure~\ref{fig5}~(solid line for MOD1 and dotted line for MOD1-bis) 
and Figure~\ref{fig6}~(dashed line), namely those with $e_1\ge 0.9$ at this instant.  
 $80\%$ of these binaries reach this point at a  GW frequency of $f_{\rm GW}\le 100$Hz.}
\end{figure}

We believe that the nature of the discrepancy  resides at least partially in the criterion used  in \citet{wen03} for
selecting stable versus unstable systems, see Equation~(42) of \citet{wen03}. This equation ignores the $(1-e_2)$ term   
that appears at the denominator in the right hand side of our Equation~(\ref{stab}).  
This could allow for initial conditions with larger values of $e_2$ and therefore smaller periapsis separation
than in our simulations and lead to an artificial bias toward large residual eccentricities
as shown in Figure~\ref{fig1}. 
We already mentioned, as an example, the system of Figure~8 of \citet{wen03}. This system
produces a very eccentric GW source, $e_1\sim0.9$ at $10~$Hz, but in reality this is a
highly unstable triple according to the \citet{MH01} stability criterion~(see also our Figure~\ref{fig1}). 
Moreover, we  note  that most of the initial conditions that in Wen's Monte-Carlo experiments would
allow the binary to enter the aLIGO band at high 
eccentricities, $a_2/a_1=3$~in her Table~1,  are unstable even according to Equation~(42) of Wen.
Additional experiments, in which we varied  integration parameters,  
changed the integrator itself, and also used a different code~\citep[that of][]{PM:11}  convinced us of the robustness of our results.

The dominant GW frequency at which  a binary  first becomes a potentially detectable GW 
source can be substantially  larger than $10$~Hz~(e.g., Figure~\ref{fig3}). This could 
 in principle affect the detectability of eccentric sources even assuming one  
has a perfect inspiral waveform template.
In fact, the detectability of a BH binary is related to
how much energy is radiated in the frequency range where aLIGO will be more 
sensitive, i.e. around 100 Hz. 
If the pericenter  of an highly eccentric orbit  is such that it has a peak frequency
above this range, then most of the GW energy would be radiated at higher frequencies
than 100 Hz, making such source  difficult to detect.
The typical GW peak frequency of eccentric sources is 
given in Figure~\ref{fig7} which displays the cumulative distribution of peak GW frequencies 
at the moment the inspiral due to GW energy loss starts. We do this only for  systems in the left panel of 
Figure~\ref{fig5} and Figure~\ref{fig6}  with residual 
eccentricities larger than $0.9$.
We find that about $80~\%$~($50~\%$)
of coalescing binaries have $f_{\rm GW}<100$~Hz~($f_{\rm GW}<40$~Hz) 
at the instant where the GW-driven inspiral begins to dominate over the Kozai-dynamics.
Thus most eccentric sources could be efficiently detected by aLIGO, provided an  efficient
search strategy.

\section{Discussion and Conclusions}\label{conc}
We have explored the dynamical evolution of hierarchical BH triples 
and investigated the conditions which  lead to the
rapid merger of  BH binaries in GCs.  We did this by using 
the direct numerical integrator AR-CHAIN which
includes  PN corrections to the equations of motion up to order 2.5 and the
algorithmically regularized chain structure  to avoid singularities~\citep{MM:08}.
We have shown that the presence of a third outer BH can  drive the inner binary at very high eccentricities; at which point
the BHs can rapidly merge before the triple system is perturbed or disrupted 
through gravitational  encounters with 
cluster field stars and on timescales much shorter than for similar binaries that are evolved in isolation.
We determined the properties of coalescing BH binaries  in GCs  and compared the results of our $3$-body
simulations to  the predictions of the standard 
 orbit average treatment in which  the equations of motion are averaged over  the rapidly varying mean 
anomalies of inner and outer orbit. 
We have demonstrated  that the orbit average treatment
leads to incorrect results  if the inner binary is orbited by an  outer perturber at a small distance~(Equation~[\ref{sp_om2}]).

The implications of our results are discussed in what follows.

\subsection{Eccentric gravitational wave sources}
The $3-$body integrations presented in this paper predict a large number~($30-50\%$) of GW sources with a substantial 
eccentricity~($e_1\gtrsim0.1$)  in the high frequency band of GW detectors. 
We  predict the existence of a large population of extremely eccentric  GW sources,~$(1-e_1)\lesssim 10^{-5}$,
that will be potentially detectable by advanced GW detectors. These mergers are driven by Kozai (like) 
dynamics in moderately hierarchical triple systems that are predicted to efficiently form via dynamical exchange
interactions  in the dens core of globular clusters.The number of these highly eccentric GW  sources is strongly 
underestimated  when adopting the secular approach.

Eccentric sources  have a  unique GW signal compared to circular binaries.
Initially, they will repeatedly burst in the aLIGO band at each periapsis approach where the GW emission is maximized.
The broad-band nature of the GW signal emitted during the repeated bursts  phase could make 
 these sources  potentially detectable  at larger distances and in a broader mass range than circular mergers~\citep[e.g.,][]{O:09,kocsis+levin12}.
For example, the inspiral of a stellar BH with another BH of  mass $\sim10^3~\msun$ may be revealed by 
aLIGO only if the event is sufficiently eccentric during plunge~\citep{O:09}.
Later  the binary transits to lower eccentricities and to a continuous powerful chirp signal within the frequency band of 
aLIGO type detectors.   

The detection of GW signal requires a bank of theoretical modeled binary inspiral waveforms as filters.
Current searches for binary BHs  rely on  \emph{circular} binary templates based
on   PN  modeling of  the inspiral phase in the weak-field slow-motion regime~\citep{blanchet06}, 
and numerical  relativity simulations to describe the merger phase in the strong gravity field regime~\citep{pretorius,campanelli,baker}.
An extensive  review of the current status of numerical simulations of compact binary mergers is given in~\citet{pfeiffer12}.
There are at least three reasons  why filter templates are currently limited to  circular binary waveforms.
First, circular binaries remain a very important GW source, probably the most important one.  
Specifically, NS-NS systems are the primary  target for aLIGO.
Second, for circular binaries, one can push the PN expansions to higher order (via energy balance arguments).  
This allows one to constract PN waveforms that are of sufficient quality for detection for total mass $m_0+m_1\leq 12~\msun$.  
Currently,  eccentric PN waveforms are not known to similar high order, thus an eccentric PN waveform of sufficient accuracy does not  exist.
Third, circular binaries tend to be dominated by (2,2) and (2,-2) radiation.  If one discards all other modes, and looks only for (2,2) 
waveform-modes, then it turns out sky-position, source-orientation, and orbital phase of the binary are all degenerate.  Therefore, the dimensionality of the parameter space dramatically collapses.  Eccentric waveforms will have a much more complex emission pattern, breaking this degeneracy.
Adding eccentricity  to binary templates would also imply that more templates are
needed which will require a higher signal to noise threshold for a given significance.  
This might reduce,  at least  in part, the detectability benefit one could  get for eccentric binaries. 

The effect of eccentricity in aLIGO searches was investigated in
\citet{brown+zimmerman10} and more recently in \citet{huerta+brown13}.
These authors find that circularized waveforms are sufficient to recover the signal
 emitted by eccentric sources  with eccentricities  less than 
$\sim 0.1$ at a fiducial GW frequency of $14~$Hz.
Thus the eccentric binaries
found in our study cannot  be efficiently detected by advanced detectors with current search pipe-lines.
Optimal searches for such systems would  require
replacement of the conventional  quasi-circular aLIGO templates with eccentric-binary templates for data analysis,
or the use of more  practical excess power searches  with  stacking~\citep{east+13}. 

Although eccentric binaries are frequent in our simulations,
we found that about half of them   will posses eccentricities smaller than~$0.1$ when their
dominant  GW frequency becomes  larger
than $40~$Hz such that they might be detected even  with regular templates. 
After detection, the careful re-analysis of the 
aLIGO data could reveal  ``pre-merger flares" from Kozai resonances.  
If those are found, the precise timing of the pre-merger flares would  carry  information about the triple system.
The detection of such systems could
give us unprecedented  insights on the dynamical processes that have
shaped the dense central environment of GCs and galaxies and  lead to  the formation and growth of   massive 
and intermediate mass BHs~\citep{miller+hamilton2002b}.

\subsection{Merger rates and black hole populations in globular clusters}
Adopting standard initial mass functions, about $1\%$ of the total mass in a stellar 
system will be in BHs that formed through  the supernova explosions 
of the most massive stars~($\gtrsim 20~M_\odot$). Soon after their formation, the BHs will be the heaviest component 
of the stellar cluster, with a mass about  $10$ times larger than the mass of a typical
 cluster star~\citep{woo+02}.
Gravitational encounters  tend to accelerate the lowest mass 
stars to the highest velocities, at the expense of the specific kinetic energy of the BHs, which   
will migrate down to the cluster core.  This process, typically referred to as ``mass-segregation''~\citep{Spitzer},
 leads to  the formation of a centrally concentrated sub-cluster 
of stellar BHs  and  to favorable conditions for the assembly of BH binaries.
The coalescence of such binaries through GW radiation could be a major observational target for the next 
generation of  ground-based GW detectors~\citep[e.g.,][]{sadowski+08}.

Due to the computational challenge of simulating the long-term evolution of massive clusters 
with direct $3$-body simulations, predictions of  event rates for the GW sources investigated in this paper remain largely uncertain.
A key question is whether  BHs  are efficiently  ejected  through
strong binary interactions  during  the cluster evolution. Efficient depletion of
the remnant population would decrease the chance for the BHs to undergo repeated exchange interactions in the cluster core
and become part of hierarchical triple systems. For the less massive and ancient GCs, the 
 BH population might self-deplete in less than a few Gyr 
before the cluster enters the cosmological  volume of  aLIGO~($z\sim2$).
This will reduce the number of systems that will manage to merge  within GCs due to the Kozai mechanism 
and become  detectable sources of GW radiation. 

Monte-Carlo simulations have been extensively  used to study the evolution of BH populations  in GCs
and to determine the event rates of BH binary mergers in the Universe~\citep[e.g.,][]{oleary+06,DBGS10,DBGS11}.
These models suggest that after  its formation the cluster's BH population typically  evaporates over a Gyr timescale.
This prediction has been  challenged by recent theoretical studies~\citep{morscher+13,sippe+hurley13} and by observational 
evidence~\citep{maccarone+07,brassington+10,maccarone+11,strader+12} which suggest
that  old GCs may still contain hundreds of stellar BHs at present. We add that most Monte-Carlo codes 
do not allow for the formation of the
long-lived hierarchical triples investigated in this paper~\citep[e.g.,][]{DBGS10,DBGS11} or when they do~\citep{oleary+06}
they employ the orbit average  formalism which we have shown to  lead to unreliable  
results and inaccurate determination of merger times~(e.g.,~Figure~\ref{fig4}). 
An isolated example of high precision $N$-body simulations of the evolution of stellar clusters containing 
a population of massive remnants was recently presented by~\citet{aarseth12}. This author finds 
that the conditions for BH binary mergers
in GCs are typically initiated by a combination of GW energy  loss and 
the Kozai resonance induced by the presence of a third object, usually another BH.

More sophisticated $N$-body simulations,
or Monte-Carlo codes including more precise prescriptions for the dynamical 
evolution of hierarchical BH triples
would be required in order to address the role of the Kozai resonance in determining the  rates
of BH binary mergers in GCs.  It is useful however to give a simple estimate of how many such eccentric events are
  likely to be detected with aLIGO. 

We estimate the detection rate of eccentric BH binary mergers due to the Kozai mechanism as:
\begin{equation}\label{rate}
 \Gamma_{\rm aLIGO}=\frac{4\pi}{3}D^3n_{\rm cl} \Gamma_{\rm merge}  f_{\rm triple} f_{\rm ecc}   
\end{equation} 
where $D$  is the maximum distance from which the emitted GW signal from a BH binary inspiral
can be detected, $n_{\rm cl}$ is the number density of GCs, $\Gamma_{\rm merg}$ is 
the merger rate of BH binaries that occur within the cluster, $f_{\rm triple}$ is the fraction of 
 such mergers induced  by the Kozai mechanism, and $f_{\rm ecc} $ is the fraction of mergers due to 
the Kozai mechanism which retain a  large eccentricity,  $e>0.1$, in the aLIGO band.
Based on the Monte Carlo simulations of~\citet{oleary+06}~(see their Table~2)  we 
take a conservative rate of $\approx 10$ mergers  inside GC  per $10^{10}$~yr
and that a fraction $ f_{\rm triple}\approx0.05$ of these mergers are due to the Kozai mechanism.
From our simulations we have $f_{\rm ecc} \approx 0.5$.
If we take  a globular number density of $10/{\rm Mpc}^3$~\citep{brodie+strader}, and that BH-BH mergers can be seen  out to
a sky-averaged distance of 1 Gpc (as expected for $10~\msun - 10~\msun$
coalescences),   and if we assume that mergers are distributed
  uniformly over the lifetimes of the globulars (as opposed to happening
  primarily early on), then Equation~(\ref{rate}) gives an aLIGO detection rate of eccentric BH binary 
  mergers due to the Kozai mechanism 
of $ \Gamma_{\rm aLIGO}\approx 1/$yr.  
Thus a uniform rate of mergers amounting to a few per Hubble time per cluster 
gives a relatively large population of eccentric BH binaries. 
We note that adopting a uniform rate of mergers might be an optimistic assumption for galaxies similar
to the Milky Way. In fact, Galactic globular clusters appear  to be
exclusively old  ($\gtrsim 10$~Gyr)  stellar systems~\citep[e.g.,][]{Ros}, implying that most BH mergers are likely 
to have occurred early on at high redshifts. 
However, this might  not be  the case in many external galaxies in which 
massive globular (like) clusters often appear to  form in an ongoing and continuous process~\citep{larsen+richtler00}. 

\subsection{Merger  of binaries with stellar components}
We finally  note  that
the results derived here could also apply to a variety of  astrophysical systems.
These include  stellar binaries,   star-compact object binaries~\citep{1979MS,2006EK}
and   binaries residing in galactic nuclei near massive BHs~\citep{ant+10,AP12,prodan+13b}. 
For these binaries GW radiation is likely not to play an important role.
Their merger  could be however  induced by the Kozai mechanism combined
with tidal friction.

These systems  could also experience the non-secular dynamical evolution discussed here, which  would 
largely increase the chance of collisions with respect to what
was previously thought.
Mergers of main sequence binaries are known to be a possible source of blue struggles in GCs~\citep{per+09c},
they might produce a population of rejuvenated  stars at the Galactic center~\citep{ant+11a}, and 
result in optical~\citep[e.g.,][]{tylenda+11} and X-ray~\citep{antonini+10b} transient events.
The formation of type Ia supernovae  could also be induced by similar processes~\citep{KD13}.
A more quantitative study of the evolution of moderately-hierarchical triples containing  binaries with main-sequence 
star components and various   types of stellar remnants will be the topic of future work.
\\

\bigskip
We thank the referee for his/her valuable comments.
We are grateful to C.~Miller, H.~Pfeiffer and S.~Prodan for useful comments that helped to improve an
earlier version of the manuscript. We acknowledge helpful discussions with Y. Lithwick and F.~Will.

\bibliographystyle{apj}

\end{document}